\documentclass[final,3p,times,twocolumn]{elsarticle}

\usepackage{amssymb}
\usepackage{amsmath}
\usepackage{multirow}
\usepackage{pifont}
\usepackage{hyperref} 
\usepackage{xcolor}

\journal{Computers in Biology and Medicine}

\begin{document}

\begin{frontmatter}



\title{Size and Smoothness Aware Adaptive Focal Loss for Small Tumor Segmentation}


\author[inst1,inst2]{Md Rakibul Islam}
\author[inst3,inst4]{Riad Hassan}
\author[inst5]{Abdullah Nazib}
\author[inst5]{Kien Nguyen}
\author[inst2]{Zahidul Islam}
\author[inst5]{Clinton Fookes}

\affiliation[inst1]{organization={Dept of CSE, Bangladesh University of Business and Technology},
            addressline={Mirpur, Dhaka}, 
            country={Bangladesh}}
\affiliation[inst2]{organization={Dept of Information and Communication Technology, Islamic University},
            addressline={Kushtia}, 
            country={Bangladesh}}
\affiliation[inst3]{organization={IICT, Bangladesh University of Engineering and Technology},
            addressline={Dhaka}, 
            country={Bangladesh}}
\affiliation[inst4]{organization={Dept of CSE, Green University of Bangladesh},
            addressline={Dhaka}, 
            country={Bangladesh}}
\affiliation[inst5]{organization={School of Electrical Engineering and Robotics},
            city={Brisbane},
            country={Australia}}

\begin{abstract}
Deep learning has achieved remarkable accuracy in medical image segmentation, particularly for larger structures with well-defined boundaries. 
However, its effectiveness can be challenged by factors such as irregular object shapes and edges, non-smooth surfaces, small target areas, etc. which complicate the ability of networks to grasp the intricate and diverse nature of anatomical regions. In response to these challenges, we propose an Adaptive Focal Loss (A-FL) that takes both object boundary smoothness and size into account, with the goal to improve segmentation performance in intricate anatomical regions.
The proposed A-FL dynamically adjusts itself based on an object's surface smoothness, size, and the class balancing parameter based on the ratio of targeted area and background. We evaluated the performance of the A-FL  on the PICAI 2022 and BraTS 2018 datasets. In the PICAI 2022 dataset, the A-FL achieved an Intersection over Union (IoU) score of 0.696 and a Dice Similarity Coefficient (DSC) of 0.769, outperforming the regular Focal Loss (FL) by 5.5\% and 5.4\% respectively. It also surpassed the best baseline by 2.0\% and 1.2\%. In the BraTS 2018 dataset, A-FL achieved an IoU score of 0.883 and a DSC score of 0.931. Our ablation experiments also show that the proposed A-FL surpasses conventional losses (this includes Dice Loss, Focal Loss, and their hybrid variants) by large margin in IoU, DSC, and other metrics. The code is available at \href{https://github.com/rakibuliuict/AFL-CIBM.git}{https://github.com/rakibuliuict/AFL-CIBM.git}.
\end{abstract}


\end{frontmatter}


\section{Introduction}
\label{sec:intro}
Precise segmentation of the disease-affected region is essential for optimal results in robotic surgery, computer-aided diagnostics, and targeted radiation therapy \cite{wu2024review}. The delineation of the target region is tedious and time consuming. Recent advancement in deep neural network has achieved remarkable success in various medical image segmentation tasks and their success particularly dependent on the shape and size of the target object. When the size of the target object is large and their boundary is relatively smooth, a model with simple architecture and trained using common loss functions is enough for the segmentation regardless of modality, organ, or lesion \cite{zhang2024segment, ronneberger2015u, roy20233d, pais20203dregnet, 10.1371/journal.pone.0304771}. Moreover, incorporating boundary priors into the loss function has been shown to improve segmentation performance \cite{liao2022lesions}.

In medical imaging, segmentation is challenging due to irregular and variable shapes of organs and tumors across different individuals. Segmentation networks often use loss functions such as Dice \cite{sudre2017generalised}, Cross-Entropy \cite{ruby2020binary}, and Focal \cite{lin2017focal}, which mainly emphasize object overlap and the entropy between pblackicted and ground truth masks. However, these loss functions frequently neglect critical factors like surface boundary characteristics and object volume, which are crucial for accurately segmenting small or irregularly shaped objects. 
Convolutional neural networks (CNN) indiscriminately extract low and high level features from both foreground and background and optimizes features based on the criteria set on the loss functions. Mere use of object overlap or entropy as optimisation objective does not give enough boost to convolutional networks to learn very small and irregular shaped tumors. 

In case of Prostate cancer, where the area of prostate gland is already small, detecting tumors in such small area of interest is a challenging task since the area of the tumor is significantly small compablack to the surrounding context. Highly non-smooth tumor boundary, therefore the irregular shape of a prostate tumor adds extra layer of challenge for the deep-learning models in their segmentation. Based on this observation,
we hypothesize that the ratio of the number of pixels presents in the foreground and background can be a discriminative signal for the network optimisation. Similarly, gradient based smoothness measure of the tumor acts as a regularizer for the network during training. Therefore, we propose size and smoothness aware adaptive focal loss (A-FL) that embeds the ratio of foreground and background (the size ratio) pixels and tumor smoothness in the conventional focal loss. Instead of a fixed class balancing parameter (\(\alpha_{va}\)), we introduce two more parameters in the focal loss setting and tested proposed A-FL on two publicly available segmentation datasets.                           
The primary contributions of this study are:
\begin{enumerate}
   \item Three adaptive parameters are introduced in the focal loss: (i) \textit{class balancing parameter} ($\mathrm{\alpha_{va}}$); (ii) \textit{volume information parameter} ($\mathrm{\gamma_{va}}$), both are calculated using combinations of foreground-background pixel ratios; and (iii) \textit{mean surface smoothness parameter} ($\mathrm{\gamma_{mSa}}$), which incorporates tumor smoothness. 
    \item A focusing parameter ($\gamma_{adaptive}$), that dynamically adjusts based on the smoothness of the segmented objects ($\gamma_{mSa}$), and volume information parameter (\(\gamma_{va}\)). This approach addresses the limitations of static parameter choices in traditional Focal Loss functions, enhancing the segmentation of irregularly shaped and small-sized tumors.
\end{enumerate}

Extensive experiments are conducted on two benchmark datasets: PICAI 2022 \cite{SAHA2024879} and BraTS 2018 \cite{menze2014multimodal}, using ResNet50\cite{he2016deep} as backbones for U-Net. The results demonstrate that A-FL outperforms the regular Focal Loss during training and testing. We also offer further insights, examining the limitations, and propose potential enhancements for A-FL through various ablation studies in the result discussion.

\section{Related Work}
Binary Cross Entropy (BCE) loss \cite{ruby2020binary} and its variations \cite{xie2015holistically, ruby2020binary, leng2022polyloss} are frequently used in semantic segmentation \cite{maninis2018deep, lin2023click}. Treating all pixels equally can lead models to focus on trivial features of easy examples while ignoring distinctive features of hard examples. To address the imbalance between positive and negative samples \cite{rahman2016optimizing, milletari2016v}, or the disparity between easy and hard samples \cite{lin2017focal, leng2022polyloss}, previous initiatives \cite{pihur2007weighted} \cite{xu2021disegnet} have proposed various adjustments.

Weighted Binary Cross Entropy (WBCE) \cite{ho2019real} introduces a weighting factor for positive samples to correct the imbalance. However, it requires manual tuning of weights, which can be tedious and risk overfitting. Balanced Cross Entropy (BCE) \cite{xie2015holistically} applies weights to both positive and negative samples, adding complexity and potentially causing instability with extremely imbalanced datasets. Both approaches attempt to mitigate BCE's bias towards majority classes but come with their own limitations. These techniques are beneficial when applied to skewed data distributions \cite{jadon2020survey}, but their impact on model performance when applied to balanced datasets may be less pronounced.  An unique solution to this problem was offeblack by Leng et al \cite{leng2022polyloss} addressed this by proposing Poly Loss (PL), which combines polynomial functions. However, they only consideblack integer powers of the polynomial terms and did not explore non-integer powers, which could offer more flexibility.

\textcolor{black}{
Dice Loss \cite{Fidon_2018} measures the overlap between pblackicted and target masks, making it effective for imbalanced datasets. However, it struggles with false positives (FP) and false negatives (FN), limiting its performance on small or irregular regions \cite{sudre2017generalised}. To address some of these issues, Jaccard Loss (IoU) \cite{rahman2016optimizing} calculates the ratio of intersection to union, which is useful for boundary delineation. Yet, it is overly sensitive to small errors, particularly in small object segmentation. In response, Tversky Loss \cite{salehi2017tversky} generalizes both Dice and Jaccard Loss by introducing hyperparameters $\alpha$  and $\beta$ offering greater flexibility in handling class imbalances and prioritizing precision or recall. However, despite its advantages, it is still sensitive to the choice of hyper-parameters, which can require extensive tuning to achieve optimal results.
}

Focal Loss (FL) \cite{lin2017focal} provides a difficulty modifier to address the discrepancy between hard/easy samples. This helps the model concentrate more on hard cases by lessening the influence of easy examples. This approach has been shown to improve segmentation performance over standard cross-entropy and Dice loss, particularly in terms of sensitivity and Dice score \cite{xu2020focal}. The Normalized Focal Loss (NFL), which incorporates an extra correction factor inversely related to the modulating component in FL, was proposed by Sofiiuk et al. \cite{sofiiuk2019adaptis}. Other research \cite{milletari2016v}, \cite{rahman2016optimizing} have also attempted to address this issue; nevertheless, difficulties such as gradient swamping impede the proper classification of equivocal pixels. These studies have used a fixed focusing parameters, particularly the study \cite{lin2017focal} have used fixed value for $\alpha$ and $\gamma$, which fails to address the varying difficulty levels in semantic segmentation, including small and irregularly shaped tumors or objects. Consequently, the model often does not achieve optimal segmentation accuracy for these challenging examples. 

We address these limitations by dynamically adjusting focusing parameter ($\gamma_{adaptive}$, and $\alpha_{va}$) based on tumor volume and surface smoothness information. The volume of a segmented object indicates its size, helping the model identify small tumors that the existing losses often miss. By considering volume, the model can focus more on small-sized objects during training. On the other hand, surface smoothness reveals the complexity of an object's shape. Irregularly shaped tumors have less smooth surfaces compablack to regular shapes. By analyzing surface smoothness, the loss function can adjust its focus on challenging examples with complex boundaries. The proposed loss function also effectively handles class imbalance, and able to give more focus on hard examples. By ensuring higher training loss for challenging examples and lower for easy ones, the proposed approach allows the model to update its weights more effectively. This approach aims to overcome the restrictions of static parameter choices encounteblack in conventional Focal Loss functions so as to improve the segmentation of irregularly shaped and small sized tumors.

\section{Methodology}
\begin{figure*}[htp]
    \centering
    \includegraphics[width=\textwidth]{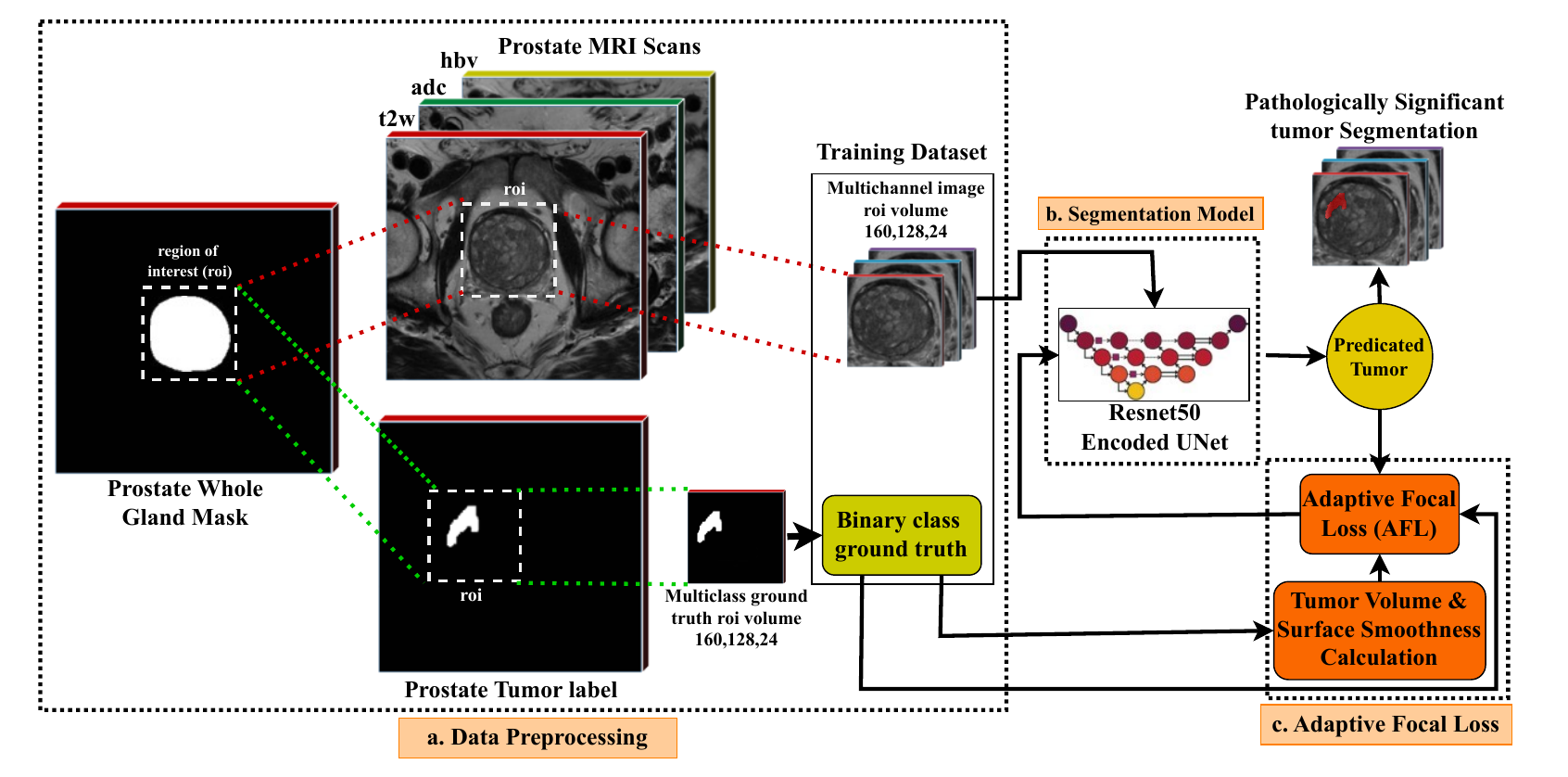}
    \caption{Overall working pipeline comprises of three main parts: a. Data pre-processing; b. Segmentation Network; c. proposed Adaptive Loss Function (A schematic overview of the training process with our Adaptive Loss function (A-FL).}
    \label{fig:OverallMethodology}
\end{figure*}

The overview of the pipeline of the proposed works is illustrated in Fig.\ref{fig:OverallMethodology}. It consists of three main components: (a) dataset pre-processing, (b) U-Net architecture, and (c) proposed A-FL loss. Details of the data pre-processing is described in Section \ref{experiment_setup}. In this section, we describe the proposed A-FL, with detail explanation of dynamically integrating tumor volume and surface smoothness information into the A-FL loss, and describe the segmentation network architecture.

\subsection{Adaptive Focal Loss}
The core concept of Adaptive Focal Loss (A-FL) is to dynamically calculate and incorporate the tumor volume and surface smoothness information into regular Focal loss function through a focusing parameter for each patient during the training process.  A-FL uses dynamically calculates the ratio of non-cancerous pixel to total pixels as class balancing parameter, which helps to address the imbalance between the numerous non-cancerous pixels and the comparatively few cancerous pixels. As shown in Fig.\ref{fig:OverallMethodology} step c, we introduce two simple but effective modifications to the regular Focal Loss function during the training process:
\begin{enumerate}
    \item During training, we assess tumor surface smoothness by computing the gradient along the x, y, and z axes, and we also evaluate tumor volume by calculating the ratio of cancerous pixels to the total pixels in the corresponding label mask. We use this smoothness and volume information as focusing parameter.
    \item We calculate the ratio of non-cancerous pixels to the total pixel count and utilize this ratio as a class balancing parameter.
\end{enumerate}

In this study, we propose an enhancement to the baseline Focal Loss by dynamically adjusting the focusing parameter during training. The following steps outline the detailed implementation of proposed Adaptive Focal Loss (A-FL):

\begin{enumerate}
    \item \textbf{Tumor Volume Based Adaptive Parameters ($\alpha_{va}$)}: To address class imbalance and give more focus to small tumor cases, we dynamically calculate the class balancing adaptive parameter and tumor volume information adaptive parameters using the cancerous and non-cancerous pixels ratios to the total pixels for each patient's tumor during training. The equations \ref{class_balancing_parameter} and \ref{volume_information_adaptive_parameter} are the mathematical formula of Class Balancing Adaptive Parameter ($\alpha_{va}$) and volume information adaptive parameter ($\gamma_{va}$).
  
        \begin{equation}
        \alpha_{va} = \frac{P_{bg}}{P_{fg} + P_{bg}}
        \label{class_balancing_parameter}
        \end{equation}

        \begin{equation}
        \gamma_{va} = \frac{P_{fg}}{P_{fg} + P_{bg}}
        \label{volume_information_adaptive_parameter}
        \end{equation}
        where \(P_{fg}\) represents the number of foreground pixels (non-zero elements) and \(P_{bg}\) represents the count of background pixels (zero elements) in the 3D mask.

    \item \textbf{Mean Surface Smoothness Adaptive Parameter ($\gamma_{mSa}$)}: To compute the mean smoothness of a patient's mask, we perform the following steps.
    \begin{enumerate}
        \item[\textit{I.}] \textit{Gradients along the x, y, and z Axes}: Let \(I\) be the image tensor. The gradients along the x, y, and z axes are denoted as \(\nabla_x I\), \(\nabla_y I\), and \(\nabla_z I\) respectively, and the formula can be expressed as in Equation \ref{gradient}.
        \begin{equation}
        \nabla_x I = \frac{\partial I}{\partial x}, \quad \nabla_y I = \frac{\partial I}{\partial y}, \quad \nabla_z I = \frac{\partial I}{\partial z}
        \label{gradient}
        \end{equation}

        \item[\textit{II.}] \textit{Gradient Magnitude}: Using the Euclidean norm of the gradients \cite{santambrogio2017euclidean} presented in Equation \ref{gradient}, we calculate the magnitude of the gradient at each point along tumor boundary.
        \begin{equation}
        \|\nabla I\| = \sqrt{(\nabla_x I)^2 + (\nabla_y I)^2 + (\nabla_z I)^2}
        \label{euclidian_norm}
        \end{equation}

        \item[\textit{III.}] \textit{Mean Smoothness}:
        The mean surface smoothness adaptive parameter (\(\gamma_{mSa}\)) is calculated as the average of the gradient magnitudes over the entire image tensor. Let \(N\) be the total number of elements in the image tensor and the formula is as follows:
        \begin{equation}
        \gamma_{mSa} = \frac{1}{N} \sum_{i=1}^{N} \|\nabla I_i\|
        \label{smoothness_calculation}
        \end{equation}
    \end{enumerate}
    \item \textbf{Adaptive Focusing Parameter ($\gamma_{adaptive}$)}:
    The adaptive parameter \(\gamma_{adaptive}\) is then calculated as the sum of the volume adaptive parameter (\(\gamma_{va}\)) and the mean smoothness adaptive parameter (\(\gamma_{mSa}\)) as follows:
    \begin{equation}
    \gamma_{adaptive} = \gamma_{va} + \gamma_{mSa}
    \end{equation}
    \item \textbf{Adaptive Focal Loss (A-FL)}: Finally, the proposed A-FL denoted as ${A-FL}(P_t)$ expands on conventional Focal loss by utilizing  dynamically adaptive parameter \(\gamma_{adaptive}\). we define $(P_t)$ as presented in Equation \ref{P_t_defination}, where p $\epsilon$ [0, 1] is the model’s estimated probability for the class with label y = 1, and y $\epsilon$ [1,0] specifies the ground-truth class.
    \begin{equation}
    p_t = 
    \begin{cases} 
    p & \text{if } y = 1 \\
    1 - p & \text{otherwise}
    \end{cases}
    \label{P_t_defination}
    \end{equation}
    The Equation \ref{a_fl} shows the mathematical formula of A-FL.
    \begin{equation}
    \text{A-FL}(P_t) = (1 - P_t)^{\gamma_{adaptive}} \cdot \log(P_t)
    \label{a_fl}
    \end{equation}
\end{enumerate}
We note two key properties of proposed Adaptive Focal Loss (A-FL):
\begin{enumerate}
    \item When an example is misclassified and  \( p_t \) is low, the modulating factor stays near 1, keeping the loss unchanged. On the other hand, as \( p_t \) nears 1, the factor blackuces to 0, thus down-weighting the loss for accurately classified examples.

    \item The value of the modulating parameter $(1 - P_t)^{\gamma_{\text{adaptive}}} $ changes in response to variations in \( p_t \) for all patients. We have investigated whether the value of \( p_t \) varies based on tumor volume and tumor surface smoothness in each patient. Thus, we incorporate these two factors (volume and smoothness) into our adaptive focusing parameter $\gamma_{\text{adaptive}}$ .
\end{enumerate}
In practice, we have incorporated the class balancing adaptive parameter $\alpha_{\text{va}}$  as defined in Equation \ref{class_balancing_parameter}, into our proposed loss function. This inclusion results in slightly better accuracy compablack to the compablack to the A-FL without $\alpha_{\text{va}}$. Thus the final A-FL loss is given by:
\begin{equation}
\textbf{A-FL}(P_t) = \alpha_{\text{va}} \cdot (1 - P_t)^{\gamma_{\text{adaptive}}} \cdot \log(P_t) \quad 
\label{final_AFL}
\end{equation}

\subsection{Segmentation Network}
\begin{figure*}[ht]
\centering
\includegraphics[width=.9\textwidth,height=0.6\textwidth]{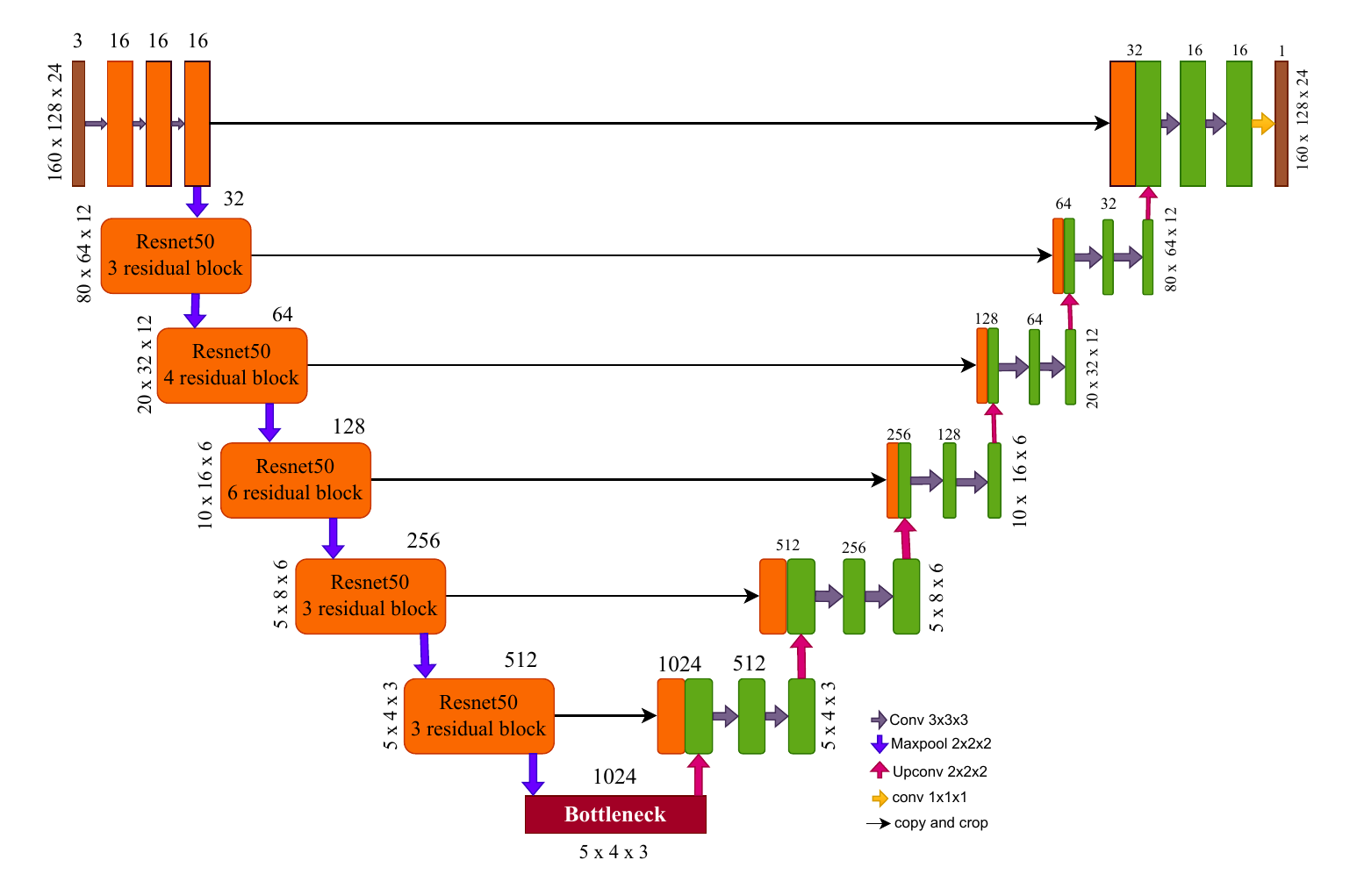}
\caption{The U-Net network architecture uses pre-trained ResNet50 \cite{he2016deep} as backbone. The residual blocks in ResNet50 enable the network to learn more complex features and deeper representations, which are crucial for accurate segmentation.} \label{resnet50_encoded_unet}
\end{figure*}

For all our experiments, we utilize ResNet50 \cite{he2016deep} pretrained weights as the encoder-backbone in the U-Net architecture. This backbone is extensively employed in semantic segmentation \cite{minaee2021image}, making it an ideal baseline for comparison and future studies. Integrating ResNet50's pretrained weights into U-Net encoder (displayed in Fig. \ref{resnet50_encoded_unet}) significantly boosts the network's feature extraction capabilities. The residual blocks in ResNet50 effectively mitigate the vanishing gradient issue, enabling the network to learn more robust and abstract features.

\section{Experiment Setup}
\label{experiment_setup}

\subsection{Dataset} All our experiments use two publicly available MRI datasets: 1) the PICAI 2022 dataset \cite{SAHA2024879} and 2) the BraTS 2018 dataset \cite{menze2014multimodal}. Both datasets are designed to improve cancer diagnosis using deep learning (DL) tools.\\
\textbf{PICAI-2022 dataset:} 
This dataset includes 1,500 Bi-parametric MRI (bpMRI) cases with three modalities: Apparent Diffusion Coefficient (ADC), High b-value (HBV), and T2-weighted (T2w). It comprises 1,075 benign or indolent prostate cancer (PCa) cases, 205 unlabeled malignant cases, and 220 manually labeled malignant cases. For training, 220 patients are utilized, divided into a training set of 180 patients (80\%) and a validation set of 40 patients (20\%). There were 2 labels with full 0's that means no lesions there, So validation set size is 38.\\
\textbf{BraTS-2018 dataset:} This dataset includes multi-modal MRI scans from 650 patients, with sequences as Fluid-Attenuated Inversion Recovery (FLAIR), T1-weighted, T1-weighted with contrast enhancement (T1ce), and T2-weighted. Of these, 484 cases are manually labeled. For model training and validation, 380 cases (80\%) are allocated for training, and 104 cases (20\%) for validation.

\subsection{Data Preparation}
To compute the prostate gland-bounding box, the whole gland mask is first binarized to separate the gland from the background. The outermost contour is identified from the binarized mask, and the largest one is selected as the gland boundary. To blackuce the risk of missing potential lesions near the prostate boundary, the bounding box is extended by 30 pixels in each direction. The prostate region is extracted from the T2W, ADC, and HBV maps using this bounding box to blackuce image and mask size for quicker experimentation. Furthermore, these extracted regions are resized to 24 × 160 × 128 for model training, while the N4 bias field correction filter is applied to the dataset to blackuce bias corruption.

To ensure uniform pixel dimensions across the four modalities in the BraTS dataset, the modalities are resampled to 1.0×1.0×1.0 mm per voxel. A two-step normalizing procedure is also implemented here, as PICAI dataset. After intensity normalization, all image modalities and labels are resized to 144 x 224 x 224 along z, x and y directions.

\subsection{Experiment Design \& Implementation}
We use the Stochastic Gradient Descent (SGD) optimizer \cite{robbins1951stochastic} for all experiments. The optimizer has a base learning rate of 0.01, a momentum of 0.9, a weight decay of 0.0001, and we train for 300 epochs. The training is conducted with a batch size of 1. In order to address the problem of over-fitting, we employ data augmentation techniques such as random affine transformations, flipping, Gaussian noise, and intensity scaling. We used PyTorch 2.3.1 on a high-performance computer configuration, comprising an Intel Xeon 2.40 GHz processor, an NVIDIA RTX 3060 GPU, and 32 GB of RAM.

\subsection{Evaluation Metrics}
Evaluation metrics are crucial for assessing the performance of segmentation models. In this study, we utilized six widely used evaluation metrics: mean Intersection over Union (IoU), Dice Similarity Coefficient (DSC), Sensitivity, Specificity, Average Surface Distance (ASD), and Hausdorff Distance (HD). Among these, IoU and DSC are region-based metrics that quantify the overlap between pblackicted and ground truth segmentations, while ASD and HD are boundary-based metrics that evaluate the accuracy of contour delineation. Sensitivity and Specificity serve as pixel-level accuracy measures, reflecting the model’s ability to correctly identify positive and negative regions, respectively. Together, these metrics provide a comprehensive framework to analyze how the proposed loss function influences segmentation performance across diverse anatomical structures.

\subsection{Comparing Methods}
To demonstrate the advancements of our proposed loss function, we integrate it into fully supervised 3D CNN-based and transformer-based methods. The evaluation includes several state-of-the-art models: 
\begin{itemize}
    \item U-Net \cite{cciccek20163d}: De-facto model for medical image segmentation. A CNN network with an encoder-decoder structure that effectively captures spatial structures. It has demonstrated strong performance on the STACOM18 LVQuan challenge dataset for accurate cardiac left ventricle (LV) segmentation \cite{kerfoot2019left}.
    
    \item Attention U-Net \cite{oktay2018attentionunetlearninglook}:A U-net variant with attention module which enhances feature representation by incorporating attention mechanisms to focus on critical regions within medical images. It has been particularly effective in segmenting pancreas structures using the CT image dataset \cite{oktay2018attention}.
    
    \item RegUNet \cite{Diakogiannis_2020}: An extension of U-Net that utilizes residual connections to improve gradient flow and mitigate vanishing gradients. It has been particularly effective in segmenting cardiac structures using the ACDC dataset \cite{zhu2024selfreg}.
    
    \item R2U-Net \cite{alom2018recurrentresidualconvolutionalneural}: A U-net variant that integrates recurrent and residual connections to capture complex spatial dependencies. It has shown excellent results in blood vessel segmentation using 3 different datasets including DRIVE, STARE, and CHASH-DB1 \cite{rajpurkar2017mura}.
    
    \item SegResNet \cite{myronenko20193d}: A robust architecture with residual blocks designed to enhance deep feature extraction and generalization. It has demonstrated strong performance on the BraTS dataset for brain tumor segmentation \cite{menze2014multimodal}.
    
    \item V-Net \cite{milletari2016v}: Specifically developed for volumetric segmentation, it employs residual connections to optimize learning in 3D medical imaging. It has demonstrated strong performance on the Prostate MRI (PROMISE12) dataset for prostate gland segmentation \cite{milletari2016v}.
    
    \item Multi-Reconstruction Recursive Residual Network (MRRN) \cite{yu2019quality}: Leverages recursive and residual mechanisms to enable multi-scale learning and refine segmentation boundaries. It has shown high performance in lung nodule segmentation tasks using the Lung Image Database Consortium (LIDC) dataset \cite{jiang2018multiple}.

    \item UNETR \cite{9706678}: A transformer-based architecture that replaces traditional convolutional encoders with a Vision Transformer (ViT) to model long-range dependencies in volumetric data. It has demonstrated strong performance on the BTCV dataset for multi-organ segmentation by effectively capturing global contextual information \cite{hatamizadeh2022unetr}. 

    \item Swin UNETR \cite{hatamizadeh2021swin}: An extension of UNETR that integrates Swin Transformers into the U-Net framework, enabling hierarchical feature learning with shifted windows. This architecture combines global context modeling with local detail preservation and has shown excellent results on 3D brain tumor segmentation in the BraTS 2021 challenge \cite{hatamizadeh2021swin}.

\end{itemize}

\section{Result and Discussion}
\label{rresult_section}

In this section, we present experiment results demonstrating the benefits of integrating a dynamic focusing parameter and adaptive imbalance weighting into the regular Focal Loss (FL). We provide quantitative comparisons between conventional FL and A-FL across various datasets. Additionally, we conduct comparative analyses of A-FL against FL using different baseline models and other losses. We also evaluate qualitative examples and perform ablation studies to assess the impact of proposed A-FL.

\subsection{Quantitative Evaluation}

Table \ref{loss_function_comparison} presents a comprehensive comparative analysis of the performance of A-FL against several baseline loss functions using a U-Net model with ResNet50 on the PICAI and BraTs datasets. The analysis reveals that A-FL consistently outperforms other loss functions across both datasets. Specifically, on the PICAI dataset, A-FL achieves an Intersection over Union (IoU) of 0.696 and a Dice coefficient (DSC) of 0.769, surpassing all other loss functions. Additionally, A-FL demonstrates a 5.5\% increase in IoU, a 5.4\% rise in DSC compablack to FL, and a 1.7\% boost in Sensitivity, showcasing its improved handling of small and irregular tumors. A slight increase in Specificity (0.05\%) further indicates balanced performance.

Similarly, on the BraTs dataset, A-FL records an IoU of 0.915 and a DSC of 0.950, indicating superior performance than regular Focal Loss. The model with A-FL shows a 4.6\% improvement in IoU and a 1.9\% increase in DSC, reflecting better segmentation accuracy. Despite a minor decrease in Sensitivity (1.2\%), the gain in Specificity (1.1\%) suggests fewer false positives and a more effective training dynamics.

In contrast, 
traditional loss functions like Cross Entropy and Tversky Loss underperform compablack to A-FL on both datasets. Cross Entropy achieves an IoU of 0.630 and a Dice coefficient (DSC) of 0.705 on PICAI, and 0.813 IoU with 0.881 DSC on BraTs. Tversky Loss slightly improves these metrics but still lags behind, with an IoU of 0.654 and a DSC of 0.726 on PICAI, and 0.875 IoU with 0.921 DSC on BraTs. IoU Loss and Dice Loss show moderate improvements, with IoUs between 0.654 and 0.665 and DSCs ranging from 0.727 to 0.739 on PICAI, and slightly better performance on BraTs. 
Even advanced loss functions like Dice Cross Entropy and Dice Focal Loss are outpaced by A-FL, despite achieving higher metrics than basic loss functions. These results highlight A-FL's superior ability to enhance segmentation accuracy and robustness across datasets.

\begin{table*}[!h]
\caption{Performance comparison of A-FL across baseline loss functions using the ResNet50 backboned UNet model on PICAI and BraTs dataset.}
\centering
\begin{tabular}{l|cccc|cccc}
\hline
\multicolumn{1}{c|}{\multirow{2}{*}{\textbf{Loss}}} & \multicolumn{4}{c|}{\textbf{PICAI Dataset}} & \multicolumn{4}{c}{\textbf{BraTs Dataset}} \\ \cline{2-9} 
\multicolumn{1}{c|}{}                                        & \multicolumn{1}{c|}{\textit{\textbf{IoU} }} & \multicolumn{1}{c|}{\textit{\textbf{Dice} }} & \multicolumn{1}{c|}{\textit{\textbf{Sens.} }} & \multicolumn{1}{c|}{\textit{\textbf{Spec.} }} & \multicolumn{1}{c|}{\textit{\textbf{IoU} }} & \multicolumn{1}{c|}{\textit{\textbf{Dice} }} & \multicolumn{1}{c|}{\textit{\textbf{Sens.} }} & \textit{\textbf{Spec.} } \\ \hline
Traversky                                               & \multicolumn{1}{c|}{0.641 }                 & \multicolumn{1}{c|}{0.654 }                  & \multicolumn{1}{c|}{0.917 }                         & \multicolumn{1}{c|}{0.948 }                          & \multicolumn{1}{c|}{0.875} & \multicolumn{1}{c|}{0.921} & \multicolumn{1}{c|}{0.5221} & 0.6499 \\ \hline
CE                                           & \multicolumn{1}{c|}{0.630 }                 & \multicolumn{1}{c|}{0.705 }                  & \multicolumn{1}{c|}{0.870 }                         & \multicolumn{1}{c|}{0.876 }                          & \multicolumn{1}{c|}{0.813} & \multicolumn{1}{c|}{0.881} & \multicolumn{1}{c|}{0.754} & 0.843 \\ \hline
IoU                                                      & \multicolumn{1}{c|}{0.654 }                 & \multicolumn{1}{c|}{0.727 }                  & \multicolumn{1}{c|}{0.917 }                         & \multicolumn{1}{c|}{0.947 }                          & \multicolumn{1}{c|}{0.893} & \multicolumn{1}{c|}{0.93} & \multicolumn{1}{c|}{0.701} & 0.796 \\ \hline
Dice                                                     & \multicolumn{1}{c|}{0.665 }                 & \multicolumn{1}{c|}{0.739 }                  & \multicolumn{1}{c|}{0.904 }                         & \multicolumn{1}{c|}{\textbf{0.952} \(\uparrow\)} & \multicolumn{1}{c|}{0.891} & \multicolumn{1}{c|}{0.933} & \multicolumn{1}{c|}{0.654} & 0.745 \\ \hline
Dice CE                                      & \multicolumn{1}{c|}{0.670 }                 & \multicolumn{1}{c|}{0.742 }                  & \multicolumn{1}{c|}{0.938 }                         & \multicolumn{1}{c|}{0.927 }                          & \multicolumn{1}{c|}{0.829} & \multicolumn{1}{c|}{0.889} & \multicolumn{1}{c|}{0.845} & 0.882 \\ \hline
DiceFocal                                              & \multicolumn{1}{c|}{0.685}                  & \multicolumn{1}{c|}{0.757 }                  & \multicolumn{1}{c|}{0.896 }                         & \multicolumn{1}{c|}{0.949 }                          & \multicolumn{1}{c|}{0.884} & \multicolumn{1}{c|}{0.927} & \multicolumn{1}{c|}{0.960} & \textbf{0.955}\(\uparrow\) \\ \hline
{FL}                                  & \multicolumn{1}{c|}{0.690}                  & \multicolumn{1}{c|}{0.726 }                  & \multicolumn{1}{c|}{0.924 }                         & \multicolumn{1}{c|}{0.947 }                          & \multicolumn{1}{c|}{0.869} & \multicolumn{1}{c|}{0.922} & \multicolumn{1}{c|}{0.954} & 0.947 \\ \hline
\textbf{A-FL}                                                         & \multicolumn{1}{c|}{\textbf{0.696} \(\uparrow\) } & \multicolumn{1}{c|}{\textbf{0.769} \(\uparrow\) } & \multicolumn{1}{c|}{\textbf{0.951} \(\uparrow\) } & \multicolumn{1}{c|}{0.948 }                          & \multicolumn{1}{c|}{\textbf{0.915} \(\uparrow\)} & \multicolumn{1}{c|}{\textbf{0.950}\(\uparrow\)} & \multicolumn{1}{c|}{\textbf{0.962}\(\uparrow\)} & {0.949} \\ \hline
\end{tabular}
\label{loss_function_comparison}
\end{table*}

\begin{table*}[!h]
\centering
\caption{Performance comparison of the A-FL function with regular FL across baseline segmentation models on the PICAI and BraTs datasets.}
\label{model_comparison}
\begin{tabular}{l|l|c|c|c|c|c|c|c|c}
\hline
\multicolumn{1}{c|}{\multirow{2}{*}{\textbf{Model}}} & \multicolumn{1}{c|}{\multirow{2}{*}{\textbf{Loss}}} & \multicolumn{4}{c|}{\textbf{PICAI Dataset}} & \multicolumn{4}{c}{\textbf{BraTs Dataset}} \\ \cline{3-10} 
\multicolumn{1}{c|}{} & \multicolumn{1}{c|}{} & \textbf{IoU} & \textbf{Dice} &  \textbf{HD} & \textbf{ASD} & \textbf{IoU} & \textbf{Dice} & \textbf{HD} & \textbf{ASD} \\ \hline

\multirow{2}{*}{UNet} & FL & 0.640 & 0.710 & 15.30 & 4.72 & 0.831 & 0.894 & 35.40 & 6.39 \\ \cline{2-10}
    & A-FL & \textbf{0.669} \(\uparrow\) & \textbf{0.740} \(\uparrow\) & \textbf{15.17} \(\downarrow\) & \textcolor{black}{\textbf{4.65}} \(\downarrow\) & \textbf{0.897}\(\uparrow\) & \textbf{0.940}\(\uparrow\) & \textcolor{black}{ \textbf{34.25}}\(\downarrow\) & \textcolor{black}{\textbf{4.11}}\(\downarrow\) \\ \hline

\multirow{2}{*}{AttUNet} & FL & 0.647 & 0.721 & \textcolor{black}{15.33}  & \textcolor{black}{{4.51}}  & 0.783 & 0.855 & \textcolor{black}{\textbf{14.63}}\(\downarrow\) & \textcolor{black}{{1.96}} \\ \cline{2-10} & A-FL & \textbf{0.679} \(\uparrow\) & \textbf{0.750} \(\uparrow\) & \textcolor{black}{\textbf{14.97}} \(\downarrow\) & \textcolor{black}{\textbf{4.33}} \(\downarrow\)  & \textbf{0.878}\(\uparrow\) & \textbf{0.926}\(\uparrow\) & \textcolor{black}{15.12} & \textcolor{black}{\textbf{1.62}}\(\downarrow\) \\ \hline

\multirow{2}{*}{RegUNet} & FL & 0.562 & 0.622 & \textcolor{black}{25.46} & \textcolor{black}{\textbf{6.81}} \(\downarrow\) & 0.722 & 0.807 & \textcolor{black}{{\textbf{42.97}}}\(\downarrow\) & \textcolor{black}{\textbf{2.51}}\(\downarrow\) \\ \cline{2-10}
                         
& A-FL & \textbf{0.573} \(\uparrow\) & \textbf{0.635} \(\uparrow\) & \textcolor{black}{\textbf{23.71}} \(\downarrow\) & \textcolor{black}{7.84} & \textbf{0.770}\(\uparrow\) & \textbf{0.841}\(\uparrow\) & \textcolor{black}{{47.72}} & \textcolor{black}{{3.60}} \\ \hline

\multirow{2}{*}{RRUNet} & FL & 0.634 & 0.709 & \textcolor{black}{14.03} & \textcolor{black}{2.99} & 0.869 & 0.992 & \textcolor{black}{30.60} & \textcolor{black}{3.49} \\ \cline{2-10}
                        
& A-FL & \textbf{0.672} \(\uparrow\) & {\textbf{0.740}} \(\uparrow\) & \textcolor{black}{\textbf{13.11}} \(\downarrow\) & \textcolor{black}{\textbf{1.47}} \(\downarrow\) & \textbf{0.883}\(\uparrow\) & \textbf{0.905}\(\uparrow\) & \textcolor{black}{\textbf{26.77}}\(\downarrow\) & \textcolor{black}{\textbf{2.18}}\(\downarrow\) \\ \hline

\multirow{2}{*}{\begin{tabular}[c]{@{}l@{}}UNet with \\ ResNet50\end{tabular}} & FL & 0.690 & 0.726 & \textcolor{black}{13.57} & \textcolor{black}{2.48} & 0.869 & 0.922 & \textcolor{black}{26.01} & \textcolor{black}{3.77} \\ \cline{2-10}
                        
& A-FL & \textbf{0.696} \(\uparrow\) & \textbf{0.769} \(\uparrow\) & \textcolor{black}{\textbf{12.97}} \(\downarrow\) & \textcolor{black}{\textbf{2.09}} \(\downarrow\) & \textbf{0.915}\(\uparrow\) & \textbf{0.950}\(\uparrow\) & \textcolor{black}{\textbf{23.91}}\(\downarrow\) & \textcolor{black}{\textbf{2.54}}\(\downarrow\) \\ \hline
\multirow{2}{*}{\textcolor{black}{SegResNet}} & \textcolor{black}{FL} & \textcolor{black}{0.626} & \textcolor{black}{0.701} & \textcolor{black}{17.41} & \textcolor{black}{2.65} & \textcolor{black}{0.849} & \textcolor{black}{0.908} & \textcolor{black}{{15.53}}  & \textcolor{black}{{2.12}}  \\ \cline{2-10}
                
&\textcolor{black}{A-FL} & \textcolor{black}{\textbf{0.691}} \(\uparrow\) & \textcolor{black}{\textbf{0.763}} \(\uparrow\) & \textcolor{black}{\textbf{15.91}} \(\downarrow\) & \textcolor{black}{\textbf{2.37}} \(\downarrow\) & \textcolor{black}{\textbf{0.889}} \(\uparrow\) & \textcolor{black}{\textbf{0.941}} \(\uparrow\) & \textcolor{black}{\textbf{12.10}}\(\downarrow\) & \textcolor{black}{\textbf{1.51}}\(\downarrow\) \\ \hline

\multirow{2}{*}{\textcolor{black}{VNeT}} & \textcolor{black}{FL} & \textcolor{black}{0.593} & \textcolor{black}{0.656} & \textcolor{black}{\textbf{17.02}} \(\downarrow\) & \textcolor{black}{{6.12}}  & \textcolor{black}{0.821} & \textcolor{black}{0.884} & \textcolor{black}{\textbf{20.52}} \(\downarrow\) & \textcolor{black}{{2.50}}  \\ \cline{2-10}
                    
& \textcolor{black}{A-FL} & \textcolor{black}{\textbf{0.626}} \(\uparrow\) & \textcolor{black}{\textbf{0.689}} \(\uparrow\) & 
\textcolor{black}{{{20.95}}}  & 
\textcolor{black}{\textbf{5.08}} \(\downarrow\) & \textcolor{black}{\textbf{0.846}} \(\uparrow\) & 
\textcolor{black}{{0.897}} &  \textcolor{black}{{23.80}}  & 
\textcolor{black}{\textbf{2.40}}\(\downarrow\) \\ \hline

\multirow{2}{*}{\textcolor{black}{MRRN}} & \textcolor{black}{FL} & \textcolor{black}{0.634} & \textcolor{black}{0.708} & \textcolor{black}{{21.48}}  & \textcolor{black}{4.22} & \textcolor{black}{0.827} & \textcolor{black}{0.889} & \textcolor{black}{21.47} & \textcolor{black}{3.18} \\ \cline{2-10}
                    
& \textcolor{black}{A-FL} & \textcolor{black}{\textbf{0.688}} \(\uparrow\) & \textcolor{black}{\textbf{0.760}} \(\uparrow\) & 
\textcolor{black}{\textbf{19.95}} \(\downarrow\)  & \textcolor{black}{\textbf{4.19}} \(\downarrow\) & 
\textcolor{black}{\textbf{0.899}} \(\uparrow\) & 
\textcolor{black}{\textbf{0.940}} \(\uparrow\) & 
\textcolor{black}{\textbf{21.19}} \(\downarrow\) & 
\textcolor{black}{\textbf{1.83}} \(\uparrow\) \\ \hline

\multirow{2}{*}{\textcolor{black}{UNETR}} & \textcolor{black}{FL} & 
\textcolor{black}{0.661} & 
\textcolor{black}{0.734} & 
\textcolor{black}{21.00}  & 
\textcolor{black}{{3.69}}  & 
\textcolor{black}{0.854}  & 
\textcolor{black}{0.912}  & 
\textcolor{black}{26.97} & 
\textcolor{black}{2.75}  \\ \cline{2-10}

& \textcolor{black}{A-FL} & 
\textcolor{black}{\textbf{0.685}} \(\uparrow\) & \textcolor{black}{\textbf{0.754}} \(\uparrow\) & 
\textcolor{black}{{\textbf{17.17}}}\(\downarrow\) & 
\textcolor{black}{{\textbf{2.75}}} \(\downarrow\)  & 
\textcolor{black}{\textbf{0.898}} \(\uparrow\) & 
\textcolor{black}{\textbf{0.941}} \(\uparrow\) & 
\textcolor{black}{\textbf{20.74}} \(\downarrow\) & 
\textcolor{black}{\textbf{2.37}} \(\downarrow\)  \\ \hline

\multirow{2}{*}{\textcolor{black}{SwinUNETR}} & \textcolor{black}{FL} & 
\textcolor{black}{0.694} & 
\textcolor{black}{0.759} & 
\textcolor{black}{{\textbf{14.11}}} \(\downarrow\)  & 
\textcolor{black}{2.51} & 
\textcolor{black}{0.913} & 
\textcolor{black}{0.937} & 
\textcolor{black}{{20.12}}  & 
\textcolor{black}{{2.23}} \\ \cline{2-10}

& \textcolor{black}{A-FL} & 
\textcolor{black}{\textbf{0.701}} \(\uparrow\) & \textcolor{black}{\textbf{0.768}} \(\uparrow\) & 
\textcolor{black}{{15.38}}  & 
\textcolor{black}{\textbf{2.45}} \(\downarrow\) & 
\textcolor{black}{\textbf{0.927}} \(\uparrow\) & 
\textcolor{black}{\textbf{0.943}} \(\uparrow\) & 
\textcolor{black}{\textbf{17.79}}\(\downarrow\)  & 
\textcolor{black}{\textbf{1.95}}\(\downarrow\)  \\ \hline

\end{tabular}
\end{table*}



Table \ref{model_comparison} highlights a consistent trend where A-FL outperforms regular FL across diverse segmentation models on both the PICAI and BraTs datasets. In models like UNet with ResNet50, A-FL delivers the highest metrics, achieving an IoU of 0.696 and a Dice coefficient of 0.769 on PICAI, along with an impressive IoU of 0.898 and a Dice coefficient of 0.951 on BraTs. This strong performance extends across multiple evaluation metrics.

Significant gains are also observed in standard CNN architectures. In the baseline UNet, A-FL raises IoU from 0.640 to 0.669 on PICAI and from 0.831 to 0.897 on BraTs, alongside Dice improvements from 0.710 to 0.740 and from 0.894 to 0.940, respectively. Similarly, AttUNet under A-FL achieves IoU and Dice of 0.679 and 0.750 on PICAI, and 0.878 and 0.926 on BraTs, consistently surpassing its FL counterpart.

The comparative analysis further expands to advanced backbones such as VNet \cite{milletari2016v}, SegResNet \cite{myronenko20193d}, and MRRN \cite{yu2019quality}. On PICAI, A-FL enhances VNet to an IoU of 0.626 and Dice of 0.689, while SegResNet reaches 0.691 IoU and 0.763 Dice, and MRRN achieves 0.688 IoU and 0.760 Dice, each reflecting stable improvements over FL. On the BraTs dataset, the pattern is even clearer: VNet improves to 0.846 IoU and 0.897 Dice, SegResNet to 0.899 IoU and 0.941 Dice, and MRRN to 0.899 IoU and 0.940 Dice.

The analysis next considers transformer-based architectures, where a subtle performance variation emerges. On the PICAI dataset, A-FL improves Dice performance, with UNETR increasing from 0.734 under FL to 0.754 and SwinUNETR rising from 0.759 to 0.768, accompanied by modest IoU gains. On the BraTs dataset, A-FL retains the advantage, as UNETR achieves a Dice of 0.941 compablack to 0.924 with FL, and SwinUNETR records 0.943 versus 0.937. Although these improvements are smaller than those observed in CNN-based backbones-likely due to the smaller number of training samples. This inspectional analysis shows that transformer models still consistently deliver strong Dice scores ($\geq 0.75$ on PICAI and $\geq 0.92$ on BraTs) and stable IoU values across both loss functions.

The surface‐based evaluation in Table \ref{model_comparison} shows that A-FL consistently improves ASD  and HD across most models, underscoring its effectiveness in refining boundary precision.

On the PICAI dataset, the most notable gain is observed in RRUNet, where ASD is blackuced by more than half from 2.99 to 1.47 (\textminus 50.8\%). Significant improvements are also seen in UNETR (\textminus 25.5\%) and UNet with ResNet50 (\textminus 15.7\%), demonstrating that A-FL substantially enhances contour accuracy in both CNN-based and transformer-based architectures. More moderate yet consistent blackuctions are evident in AttUNet (4.51 to 4.33) and SegResNet (2.65 to 2.37), reinforcing the stability of the proposed loss. A few exceptions emerge in RegUNet, where ASD increases from 6.81 to 7.84, and in VNet, which shows inconsistent boundary improvements. Inspectionally, both RegUNet and VNet also exhibit comparatively lower Dice and IoU scores, suggesting that limited volumetric accuracy constrains their ability to achieve consistent surface refinement. These ASD patterns are further supported by blackuctions in boundary errors measublack by HD, with notable improvements in models such as UNETR (\textminus 18.2\%) and SegResNet (\textminus 8.6\%).

On the BraTs dataset, the improvements in ASD are even more pronounced, with several models showing large relative gains under A-FL. The strongest refinement is achieved by MRRN, where ASD decreases from 3.18 to 1.83 (\textminus 42.5\%), followed closely by SwinUNETR, which blackuces ASD from 2.75 to 1.95 (\textminus 29.1\%), and SegResNet, which improves (\textminus28.8\%). UNet also demonstrates a substantial blackuction, lowering ASD from 6.39 to 4.11 (\textminus 35.7\%), confirming that A-FL is highly effective for conventional CNN backbones. Transformer-based architectures further reinforce this trend: UNETR lowers ASD (\textminus 13.8\%) while also blackucing HD (\textminus 23.1\%), and SwinUNETR achieves simultaneous improvements in HD (\textminus 11.6\%). In contrast, a few cases show marginal degradation, for example UNet with ResNet50 where ASD rises slightly from 2.54 to 3.77. Collectively, these results suggest that although A-FL consistently strengthens surface accuracy, the extent of improvement is influenced by architectural characteristics, thereby highlighting its architecture-sensitive nature.

Tables  \ref{loss_function_comparison}, and \ref{model_comparison} provide evidence of the superiority of A-FL in segmentation tasks, demonstrating significant improvements in IoU, Dice, ASD, HD, sensitivity and specificity across different datasets, loss functions, and models. These results underscore A-FL's effectiveness in addressing class imbalance through the use of $\alpha_{\text{va}}$, calculated using Eq. \ref{class_balancing_parameter}, thereby enhancing performance in segmenting small and irregular tumor volumes.

\begin{figure}[!h]
\centering
\includegraphics[width=0.47\textwidth]{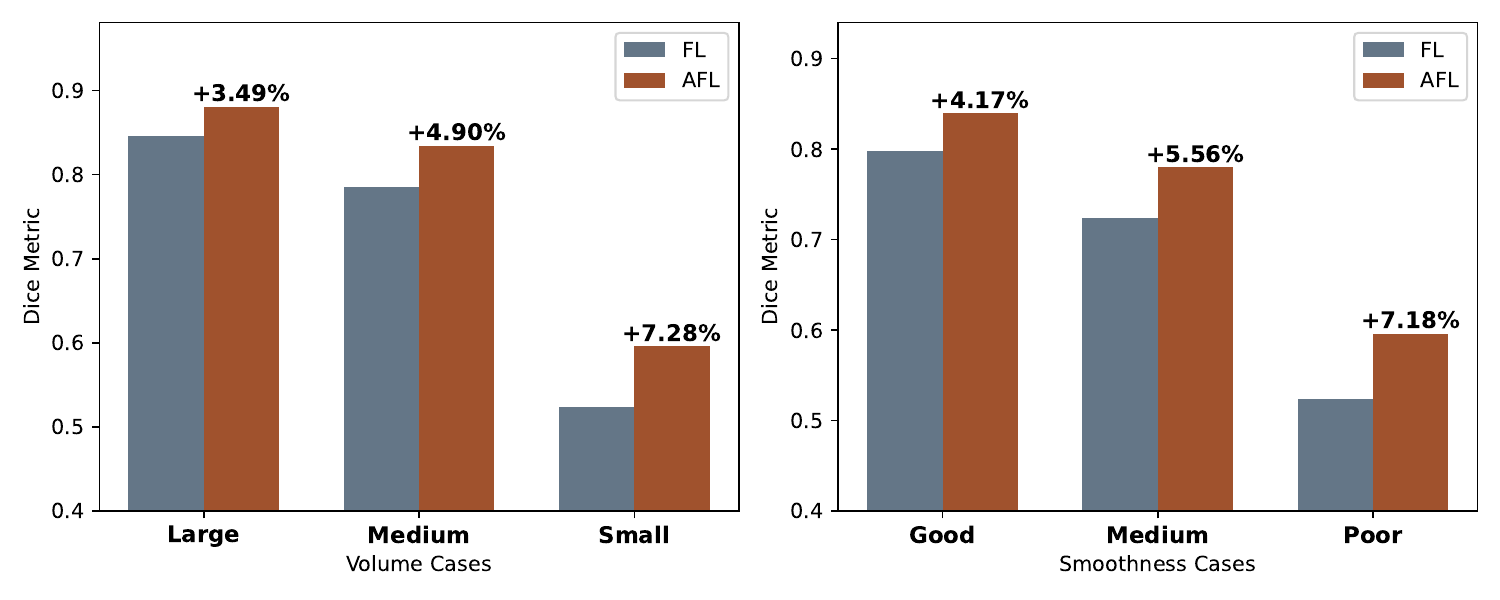}
\caption{Performance comparison of regular Focal Loss (FL) vs. Adaptive Focal Loss (A-FL) on PICAI dataset.
The left chart shows average Dice Similarity Coefficient (DSC) for large, medium, and small volume cases.The right chart shows average DSC for good, medium, and poor smoothness cases.  Results highlight A-FL's effectiveness in improving segmentation accuracy, especially for small volume and irregularly shaped tumors.}
\label{comparison_chart}
\end{figure}

\subsection{Effect of A-FL on Varying Size and Smoothness}
The bar charts in Fig. \ref{comparison_chart} compare traditional FL and the proposed A-FL using the Dice Similarity Coefficient (DSC) across different tumor smoothness and volume levels. Tumors were classified by smoothness based on surface gradient magnitudes: poor (1–150), medium (151–400), and good (401 and above). We have consideblack five cases from each group and calculate the average DSC for those cases from A-FL and FL section. A-FL demonstrates clear superiority in these categories, with a 4.17\% improvement in DSC for tumors with good smoothness, a 5.56\% improvement for medium smoothness tumors, and a notable 7.18\% increase for tumors with poor smoothness. This highlights A-FL's exceptional ability to manage tumors with varying surface complexities, particularly those with poor smoothness.

For volume, tumors were classified as small (0.1-40 mm\(^3\)), medium (41-99 mm\(^3\)), and large (101 mm\(^3\) and above) cases. A-FL shows robust performance across these categories, providing a 3.49\% improvement in DSC for large volume tumors, a 4.90\% increase for medium volume tumors, and an impressive 7.28\% boost for small volume tumors. These findings underscore A-FL's effectiveness in accurately segmenting tumors of different sizes, excelling especially with smaller, more challenging tumors.

\begin{figure}[!h]
\centering
\includegraphics[width= 0.47\textwidth]{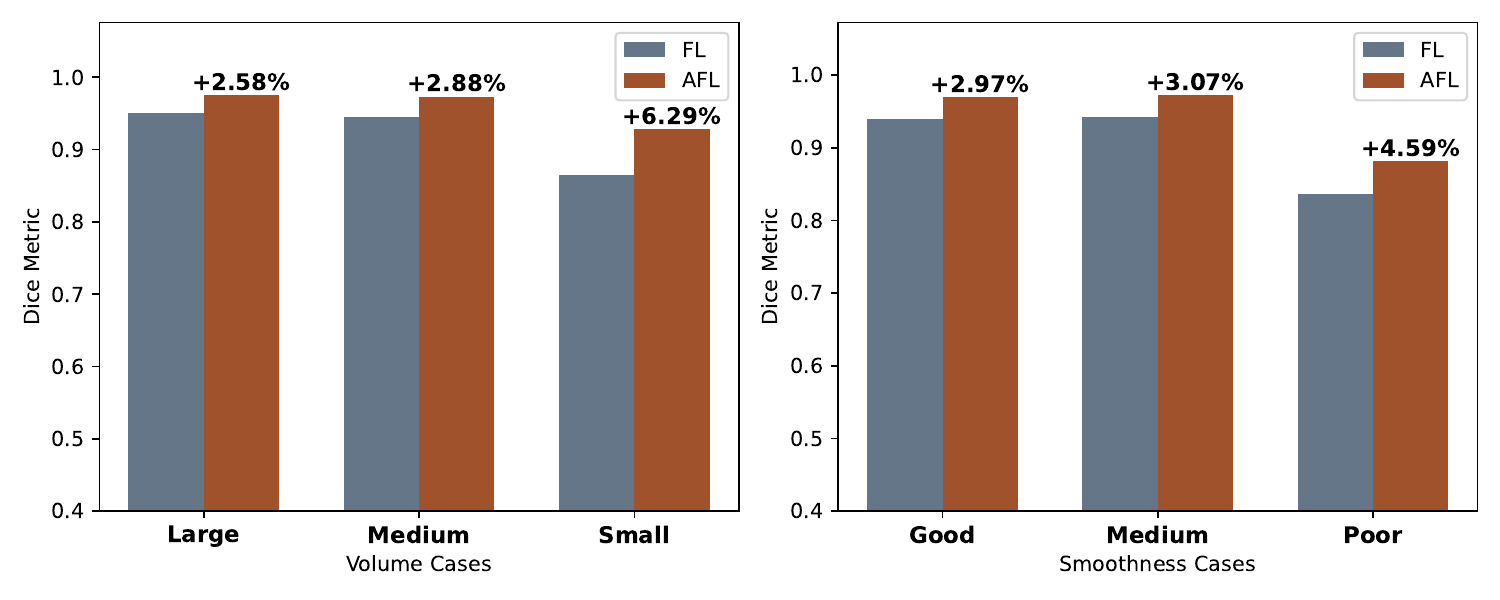}
\caption{Performance comparison of regular Focal Loss (FL) vs. Adaptive Focal Loss (A-FL) on BraTs dataset.
The left chart shows average Dice Similarity Coefficient (DSC) for large, medium, and small volume cases.The right chart shows average DSC for good, medium, and poor smoothness cases.  Results highlight A-FL's effectiveness in improving segmentation accuracy, especially for small volume and irregularly shaped tumors.}
\label{BraTS_comparison_chart}
\end{figure}

The bar charts in Fig. \ref{BraTS_comparison_chart} compare the traditional Focal Loss (FL) and the proposed Adaptive Focal Loss (A-FL) based on the DSC across different tumor volume and smoothness categories for BraTs dataset. Tumors were categorized by volume into large, medium, and small cases. AFL consistently outperforms FL across these categories, demonstrating a significant improvement in DSC, with a 2.58\% increase for large volume tumors, a 2.88\% increase for medium volume tumors, and a 6.29\% increase for small volume tumors. This indicates that AFL is particularly effective at handling small tumors, which are typically more challenging to segment. Similarly, tumors were categorized by surface smoothness into good, medium, and poor cases. A-FL again shows superior performance, with a 2.97\% improvement in DSC for tumors with good smoothness, a 3.07\% improvement for medium smoothness tumors, and a substantial 4.59\% increase for tumors with poor smoothness.

From Figures \ref{comparison_chart} and \ref{BraTS_comparison_chart}, it is evident that A-FL significantly outperforms regular FL, especially in challenging scenarios like small tumors and those with poor surface smoothness. The proposed loss function demonstrates a distinct advantage, excelling more in small tumors compablack to medium and large ones, and in tumors with poor smoothness over those with medium and good smoothness. The greater increase in DSC observed in Figure \ref{comparison_chart} compablack to Figure \ref{BraTS_comparison_chart} can be attributed to the more complex nature of the PICAI dataset, where the prostate zones are smaller, and the tumors are not only smaller but also exhibit more irregular, zigzag patterns compablack to the brain tumors in the BraTS dataset.

\subsection{Qualitative Evaluation}

\begin{figure*}[ht]
    \centering
    \includegraphics[width=.8\textwidth,height=0.6\textwidth]{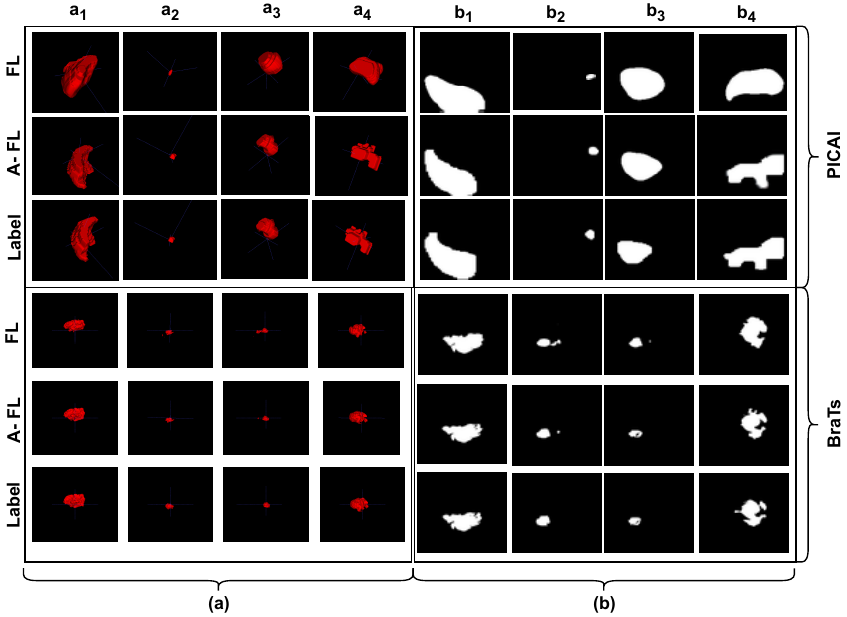}
    \caption{Qualitative results for the PICAI and BraTs validation sets are shown in 3D (\textbf{a}) and 2D slice views (\textbf{b}). The \textit{Label} row shows ground truth, the \textit{A-FL} row shows pblackictions from our method, and the \textit{FL} row shows baseline Focal Loss pblackictions. For PICAI, \textbf{(a)} presents large (a\textsubscript{1}) and small (a\textsubscript{2}) volumes, along with good (a\textsubscript{3}) and poor (a\textsubscript{4}) smoothness. In \textbf{(b)}, the $14^{th}$ and $2^{nd}$ slices show large (b\textsubscript{1}) and small (b\textsubscript{2}) volumes, while the $9^{th}$ and $7^{th}$ slices show good (b\textsubscript{3}) and poor (b\textsubscript{4}) smoothness. For BraTs, \textbf{(a)} similarly shows large and small volumes, with good and poor smoothness. In \textbf{(b)}, the $111^{th}$ and $57^{th}$ slices depict large (b\textsubscript{1}) and small (b\textsubscript{2}) volumes, and the $50^{th}$ and $94^{th}$ slices show good (b\textsubscript{3}) and poor (b\textsubscript{4}) smoothness.}
    \label{2D_3D_comparison}
\end{figure*}

The qualitative results for the PICAI dataset, shown in Fig. \ref{2D_3D_comparison}, illustrate that A-FL outperforms baseline FL in prostate cancer (PCa) segmentation. While for large volume and good smooth surface cases (a\textsubscript{1}) and (a\textsubscript{3}) show similar performance for both methods, A-FL excels in segmenting uneven surface shaped or small volume tumors (cases (a\textsubscript{2}) and (a\textsubscript{4})) as illustrated in the 3D visualizations in Fig. \ref{2D_3D_comparison}a. 

The significance of A-FL is more evident when observing Fig. \ref{2D_3D_comparison}b. Column b\textsubscript{1} displays the $14^{th}$ slice, representing the largest volume case (115 mm\(^3\)). Here, the A-FL output closely matches the label image, while the FL output is less accurate. Column b\textsubscript{2} shows the $2^{nd}$ slice of the smallest volume case (14.4 mm\(^3\)), where A-FL again produces a slice that is almost identical to the label, demonstrating its superior accuracy in low-volume cases. In both instances, the volumes are calculated using Eq. \ref{volume_information_adaptive_parameter}. In Column b\textsubscript{3}, which presents the $9^{th}$ slice with the highest smoothness (score 715), A-FL continues to deliver results that are more similar to the label image than those from FL. Lastly, Column b\textsubscript{4} illustrates the $7^{th}$ slice with the worst smoothness (score 12), where A-FL maintains a closer resemblance to the label image compablack to FL, highlighting its ability to generate more accurate results even in cases with poor smoothness.  Both smoothness scores are calculated using Eq. \ref{smoothness_calculation}.

\subsection{The Ablation Study}
To assess the effectiveness of proposed Adaptive Focal Loss (A-FL), twelve different experiments are conducted and presented on Table-\ref{ablation_study}. Each variable and combination of variables is tested using ResNet50 with U-Net architecture. The accuracy is measublack by mean IoU and mean Dice from the respective dataset.

In Table \ref{ablation_study}, the first row is the baseline results using the original Focal Loss (FL) in where $\alpha$ and $\gamma$ is manually set to 0.25 and 2 respectively \cite{lin2017focal}. while the 2nd row shows the results after
introducing $\alpha_{\text{va}}$ alone which leads to a slight improvements due to assigning higher weights to minority foreground class. On the PICAI dataset, IoU increased by 1.3\% and Dice by 1.8\%, while on the BraTS dataset, IoU increased by 0.7\% and Dice by 1.3\%. 

Combining the background volume adapting weight $\alpha_{\text{va}}$ with the foreground volume adapting weight $\gamma_{\text{va}}$ improves results by balancing both background and foreground adjustments, leading to more precise segmentation. This increased IoU by 3.6\% and Dice by 3.3\% on the PICAI dataset, and IoU by 1.5\% and Dice by 1.9\% on the BraTS dataset. Similarly, combining $\alpha_{\text{va}}$ with the most influential parameter of our A-FL, tumor surface mean smoothness adapting weight $\gamma_{\text{mSa}}$ enhances performance by ensuring more focus regarding more zigzag shaped tumor, increasing IoU by 4.6\% and Dice by 4.1\% on the PICAI dataset, and IoU by 2.6\% and Dice by 3.6\% on the BraTS dataset.

The optimal configuration is achieved when all three parameters are enabled (shows in row 4), resulting in the highest metrics. Compablack to the baseline, IoU on the PICAI dataset increased by 5.5\% and Dice by 5.4\%, while on the BraTS dataset, IoU increased by 5.2\% and Dice by 4.8\%. Conversely, disabling $\alpha_{\text{va}}$ while enabling both $\gamma_{\text{va}}$ and $\gamma_{\text{mSa}}$ results in lower performance compablack to the optimal configuration. This demonstrates the effectiveness of dynamic parameter adjustments in addressing class imbalance and complex tumor structures.

\begin{table}[ht]
\caption{Ablation studies on PICAI and BraTS dataset. Here, the values of $\alpha_{\text{va}}$, $\gamma_{\text{va}}$ and $\gamma_{\text{mSa}}$ are continuously changed for every patient in the training session.}
\centering
\begin{minipage}{0.45\textwidth} 
\centering
\small
\begin{tabular}{c|c|c|cc|cc}
\hline
\multicolumn{3}{c|}{\textbf{Parameters}} & \multicolumn{2}{c|}{\textbf{PICAI Dataset}} & \multicolumn{2}{c}{\textbf{BraTS Dataset}} \\ \cline{1-7} 
\textbf{$\alpha_{\text{va}}$} & \textbf{$\gamma_{\text{va}}$} & \textbf{$\gamma_{\text{mSa}}$} & \multicolumn{1}{c|}{\textit{\textbf{IoU} }} & \multicolumn{1}{c|}{\textit{\textbf{Dice} }} & \multicolumn{1}{c|}{\textit{\textbf{IoU} }} & \textit{\textbf{Dice} } \\ \hline
\ding{55} & \ding{55} & \ding{55} & \multicolumn{1}{c|}{0.641} & \multicolumn{1}{c|}{0.715} & \multicolumn{1}{c|}{0.831} & 0.883 \\ \hline
\ding{51} & \ding{55} & \ding{55} & \multicolumn{1}{c|}{0.656} & \multicolumn{1}{c|}{0.733} & \multicolumn{1}{c|}{0.838} & 0.906 \\ \hline
\ding{51} & \ding{51} & \ding{55} & \multicolumn{1}{c|}{0.677} & \multicolumn{1}{c|}{0.748} & \multicolumn{1}{c|}{0.846} & 0.912 \\ \hline
\ding{51} & \ding{55} & \ding{51} & \multicolumn{1}{c|}{0.687} & \multicolumn{1}{c|}{0.756} & \multicolumn{1}{c|}{0.857} & 0.929 \\ \hline

\ding{55} & \ding{51} & \ding{51} & \multicolumn{1}{c|}{0.677 } & \multicolumn{1}{c|}{0.746 } & \multicolumn{1}{c|}{0.882 } & 0.924  \\ \hline
\ding{51} & \ding{51} & \ding{51} & \multicolumn{1}{c|}{\textbf{0.696} \(\uparrow\) } & \multicolumn{1}{c|}{\textbf{0.769} \(\uparrow\) } & \multicolumn{1}{c|}{\textbf{0.915} \(\uparrow\) } & \textbf{0.950} \(\uparrow\) \\ \hline
\end{tabular}
\end{minipage}
\label{ablation_study}
\end{table}

\begin{table*}[ht]
\centering
\caption{Comparison of Performance Metrics for Various Loss Functions on the PICAI Dataset (38 Patients), Including Computation Time in millisecond (ms), Confidence Intervals, Statistical Significance (p-values), Lesion-wise Classification Accuracy Metrics (TP, FP, FN), Precision, and Recall}
\begin{tabular}{|c|c|c|c|c|c|c|c|c|} 
\hline
\textcolor{black}{\textbf{Loss}} & \textcolor{black}{\textbf{Time (ms)}} & \textcolor{black}{\textbf{Confidence Interval}} & \textcolor{black}{\textbf{P-Value}} & \textcolor{black}{\textbf{TP}} & \textcolor{black}{\textbf{FP}} & \textcolor{black}{\textbf{FN}} & \textcolor{black}{\textbf{Precision}} & \textcolor{black}{\textbf{Recall}} \\ \hline
\textcolor{black}{Tversky} & \textcolor{black}{188.84} & \textcolor{black}{(0.64, 0.64)} & \textcolor{black}{$<$0.0001} & \textcolor{black}{24} & \textcolor{black}{27} & \textcolor{black}{15} & \textcolor{black}{0.471} & \textcolor{black}{0.600} \\ \hline
\textcolor{black}{CE} & \textcolor{black}{189.58} & \textcolor{black}{(0.65, 0.69)} & \textcolor{black}{$<$0.0001} & \textcolor{black}{30} & \textcolor{black}{16} & \textcolor{black}{9} & \textcolor{black}{0.652} & \textcolor{black}{0.769} \\ \hline
\textcolor{black}{IoU} & \textcolor{black}{189.91} & \textcolor{black}{(0.68, 0.71)} & \textcolor{black}{$<$0.0001} & \textcolor{black}{34} & \textcolor{black}{32} & \textcolor{black}{5} & \textcolor{black}{0.515} & \textcolor{black}{0.739} \\ \hline
\textcolor{black}{Dice} & \textcolor{black}{188.73} & \textcolor{black}{(0.70, 0.72)} & \textcolor{black}{$<$0.0001} & \textcolor{black}{33} & \textcolor{black}{20} & \textcolor{black}{6} & \textcolor{black}{0.623} & \textcolor{black}{0.846} \\ \hline
\textcolor{black}{DiceCE} & \textcolor{black}{189.62} & \textcolor{black}{(0.73, 0.74)} & \textcolor{black}{$<$0.0001} & \textcolor{black}{28} & \textcolor{black}{24} & \textcolor{black}{11} & \textcolor{black}{0.586} & \textcolor{black}{0.773} \\ \hline
\textcolor{black}{DiceFocal} & \textcolor{black}{189.82} & \textcolor{black}{(0.73, 0.74)} & \textcolor{black}{$<$0.0001} & \textcolor{black}{32} & \textcolor{black}{13} & \textcolor{black}{7} & \textcolor{black}{0.500} & \textcolor{black}{0.706} \\ \hline
\textcolor{black}{FL} & \textcolor{black}{\textbf{183.34}} & \textcolor{black}{(0.68, 0.71)} & \textcolor{black}{$<$0.0001} & \textcolor{black}{34} & \textcolor{black}{18} & \textcolor{black}{5} & \textcolor{black}{0.507} & \textcolor{black}{0.872} \\ \hline
\textcolor{black}{AFL} & \textcolor{black}{198.00} & \textcolor{black}{\textbf{(0.74, 0.76)}} & \textcolor{black}{-} & \textcolor{black}{35} & \textcolor{black}{12} & \textcolor{black}{4} & \textcolor{black}{\textbf{0.745}} & \textcolor{black}{\textbf{0.897}} \\ \hline
\end{tabular}
\label{PICAI_results}
\end{table*}

\begin{table*}[ht]
\centering
\caption{Comparison of Performance Metrics for Various Loss Functions on the BraTS Dataset (104 Patients), Including Computation Time in milliseconds (ms), Confidence Intervals, Statistical Significance (p-values), Lesion-wise Classification Accuracy Metrics (TP, FP, FN), Precision, and Recall}
\begin{tabular}{|c|c|c|c|c|c|c|c|c|}
\hline
\textcolor{black}{\textbf{Loss}} & \textcolor{black}{\textbf{Time(ms)}} & \textcolor{black}{\textbf{Confidence Interval}} & \textcolor{black}{\textbf{P-Value}} & \textcolor{black}{\textbf{TP}} & \textcolor{black}{\textbf{FP}} & \textcolor{black}{\textbf{FN}} & \textcolor{black}{\textbf{Precision}} & \textcolor{black}{\textbf{Recall}} \\ \hline
\textcolor{black}{Tversky} & \textcolor{black}{191.91} & \textcolor{black}{(0.85, 0.90)} & \textcolor{black}{$<$0.0001} & \textcolor{black}{170} & \textcolor{black}{416} & \textcolor{black}{43} & \textcolor{black}{0.290} & \textcolor{black}{0.798} \\ \hline
\textcolor{black}{CE} & \textcolor{black}{190.70} & \textcolor{black}{(0.86, 0.87)} & \textcolor{black}{0.0012} & \textcolor{black}{143} & \textcolor{black}{550} & \textcolor{black}{70} & \textcolor{black}{0.206} & \textcolor{black}{0.671} \\ \hline
\textcolor{black}{IoU} & \textcolor{black}{190.64} & \textcolor{black}{(0.89, 0.92)} & \textcolor{black}{0.0029} & \textcolor{black}{172} & \textcolor{black}{158} & \textcolor{black}{41} & \textcolor{black}{0.521} & \textcolor{black}{\textbf{0.807}} \\ \hline
\textcolor{black}{Dice} & \textcolor{black}{\textbf{189.91}} & \textcolor{black}{(0.88, 0.92)} & \textcolor{black}{$<$0.0001} & \textcolor{black}{168} & \textcolor{black}{233} & \textcolor{black}{45} & \textcolor{black}{0.419} & \textcolor{black}{0.789} \\ \hline
\textcolor{black}{DiceCE} & \textcolor{black}{190.07} & \textcolor{black}{(0.87, 0.88)} & \textcolor{black}{$<$0.0001} & \textcolor{black}{126} & \textcolor{black}{116} & \textcolor{black}{87} & \textcolor{black}{0.521} & \textcolor{black}{0.592} \\ \hline
\textcolor{black}{DiceFocal} & \textcolor{black}{193.09} & \textcolor{black}{(0.87, 0.91)} & \textcolor{black}{0.0014} & \textcolor{black}{156} & \textcolor{black}{129} & \textcolor{black}{57} & \textcolor{black}{0.547} & \textcolor{black}{0.732} \\ \hline
\textcolor{black}{FL} & \textcolor{black}{191.87} & \textcolor{black}{(0.89, 0.91)} & \textcolor{black}{$<$0.0001} & \textcolor{black}{145} & \textcolor{black}{178} & \textcolor{black}{68} & \textcolor{black}{0.449} & \textcolor{black}{0.681} \\ \hline
\textcolor{black}{AFL} & \textcolor{black}{201.30} & \textcolor{black}{\textbf{(0.93, 0.94)}} & \textcolor{black}{-} & \textcolor{black}{158} & \textcolor{black}{116} & \textcolor{black}{55} & \textcolor{black}{\textbf{0.577}} & \textcolor{black}{0.741} \\ \hline
\end{tabular}
\label{BraTS_results}
\end{table*}

\subsection{Statistical Analysis}

It is a common practice \cite{Pellicer-Valero2022},\cite{Hong2023-pe},\cite{Rodrigues2023} to calculate statistical significance for medical image segmentation methods. The P-values and confidence-intervals are essential metrics to validate the performance of segmentation models to ensure that performance is not coming from random fluctuations but actual improvement. They also ensures the clinical reliability of the method by statistically comparing pblackictions with expert annotations.
To ensure a comprehensive evaluation of our proposed loss function, we conducted an extensive computational cost analysis along with statistical assessments, including confidence intervals, p-values, and diagnostic metrics such as false positives (FP), false negatives (FN), and true positives (TP).
These analyses provide a deeper understanding of the model's reliability, robustness, and statistical significance. The results which offer crucial insights into the performance variations and generalizability of our approach, are presented in Tables \ref{PICAI_results}, \ref{BraTS_results} for the PICAI and BraTS datasets respectively.

\textcolor{black}{
On the PICAI validation set, which includes 39 tumors across 38 patients, AFL achieves only 12 FP and significantly blackuces FN to just 4, outperforming Tversky Loss with 27 FP and 16 FN, and Focal Loss with 5 FN. AFL also achieves high precision and recall values of 0.745 and 0.897, respectively. Similarly, on the BraTS validation set, comprising 213 tumors in 104 patients, AFL records just 116 FP and 55 FN, again outperforming Tversky Loss with 416 FP and Dice Loss with 233 FP, while achieving precision and recall of 0.577 and 0.742. Across both datasets, AFL demonstrates a balanced performance with fewer errors in false positives and false negatives compablack to other loss functions.
}
\textcolor{black}{
We conducted statistical assessment of dice scores using the Wilcoxon signed-rank test across the PICAI and BraTS datasets, AFL demonstrated a statistically significant performance improvement, with a minimal confidence interval of (0.74, 0.76) on PICAI and (0.93, 0.94) on BraTS, accompanied by consistently low p-values. These findings, along with similarly low p-values observed across other loss functions, confirm AFL’s high precision in segmenting small and irregular tumors while maintaining statistical robustness. The results also indicate that AFL outperforms other loss functions in both accuracy and reliability across diverse datasets.
}
\subsection{Computational Cost}
\textcolor{black}{
We have calculated the computational costs of each loss function and the results are presented in Table \ref{PICAI_results} and \ref{BraTS_results} for both datasets. Due to AFL’s adaptive nature, which dynamically tunes both focusing ($\gamma_{adaptive}$) and class-balancing($\alpha_{va}$)  parameters, it incurs a slightly higher computational cost compablack to other loss functions. On the PICAI dataset, AFL requires an average time of 198.00 ms per patient, while on the BraTS dataset, the time increases to 201.30 ms. AFL’s computational cost is approximately 8.0\% more than Focal Loss on PICAI and 4.9\% more on BraTS, and about 4.9\% more time than Dice Loss on PICAI and 6.0\% more on BraTS. Simpler loss functions process data more rapidly per patient, though they lack AFL’s tailoblack adaptability, which is essential for capturing the small and zigzag-shaped variability of tumor shape in 3D. Despite AFL's computational expense, its high TP rate,low FP rate and confidence interval reveals true significance and indicates it potential in clinical setting.       
}

\section{Conclusion}
This paper introduces a loss function named as Adaptive Focal Loss (A-FL) tailoblack for semantic segmentation, specifically addressing tumor volume and surface smoothness considerations. A-FL improves upon traditional Focal Loss by dynamically adjusting focusing and balancing parameters at the pixel level during training. This adaptation allows our models to achieve more balanced and precise segmentation performance by integrating tumor volume and surface smoothness as focal parameters, while also considering background volume for class balancing. Experimental evaluations conducted on the PICAI and BraTS datasets using ResNet50-based U-Net architecture demonstrate the superior performance of A-FL compablack to conventional Focal Loss methods. \textcolor{black}{Despite the proven capacity, the  A-FL has so far been developed and evaluated primarily for binary segmentation tasks, and its capacity and applicability for multi-class or multi-organ segmentation is still untested. Moreover, the current experiments were confined to specific datasets with hyperparameters tuned accordingly, underscoring the need to extend the approach to multi-dataset composed training scenarios. As part of future research, we intend to investigate adaptive hyperparameter strategies and expand the applicability of A-FL to multi-class and multi-organ segmentation tasks.}


\section{CblackiT authorship contribution statement}
\textbf{MD Rakibul Islam}: Conceptualization, Methodology, Experimentation, Formal analysis, Coding, Writing - Original draft. 
\textbf{Riad Hassan}: Conceptualization, Formal analysis, Experimentation,
\textbf{Abdullah Nazib}: Conceptualization, Methodology, Writing, Project Supervision.
\textbf{Kien Nguyen}: Writing review and editing.
\textbf{Zahidul Islam}: Writing review and editing, Project Supervision.
\textbf{Clinton Fookes}: Project Supervision, Collaboration.

\section{Acknowledgments}
 The authors thank the Computer Vision \& Intelligent Interfacing (CVIIL) Lab, Islamic University, Bangladesh for providing computational resources for experiments. Special thanks to \href{https://biorainlab.net}{Biomedical Robotics, AI, and Imaging Network (BioRain) Lab} and SAIVT, QUT for their ongoing support, contributions to the study design, and project supervision.

 \bibliographystyle{elsarticle-num} 
 \bibliography{Ref.bib}





\end{document}